\documentstyle[12pt,prc,preprint,tighten,aps]{revtex}
\begin {document}
\draft

\title{
On the Coulomb Sturmian matrix elements of relativistic 
Coulomb Green's operators
}

\author{B. K\'onya and Z. Papp}

\address{
Institute of Nuclear Research of the Hungarian
Academy of Sciences, \\
P.\ O.\ Box 51, H--4001 Debrecen, Hungary 
}

\maketitle

\begin{abstract}
The  Hamiltonian of the radial Coulomb Klein-Gordon and second order
Dirac equations are shown to possess an infinite symmetric tridiagonal matrix 
structure on the relativistic Coulomb Sturmian basis. 
This allows  us to 
give  an analytic representation for  the corresponding
Coulomb Green's operators in terms of continued fractions. 
The poles of the Green's matrix reproduce the exact relativistic
hydrogen spectrum.
\end{abstract}

\vspace*{2cm}
%\pacs{PACS numbers: }

\section{Introduction}
In quantum mechanics the knowledge of the
Green's operator is equivalent to the complete knowledge of the system.
So, having an analytic basis representation for the Green's operator 
can tremendously simplify the actual calculations. 
If we know the Green's operator only of the asymptotic part of the
Hamiltonian we can treat the remaining terms as perturbations and approximate 
them by finite matrices.

In a recent publication, Ref.\ \cite{klp1}, we have proposed a method
for calculating matrix representation of Green's operators. 
If, in some basis representation, the Hamiltonian possesses an infinite 
symmetric tridiagonal (Jacobi) matrix structure the corresponding Green's 
operator can be 
given in terms of continued fractions. In Ref.\ \cite{klp1}, this theorem
were exemplified with the Green's operator of the 
non-relativistic Coulomb and harmonic oscillator Hamiltonian, and, 
in Ref.\ \cite{klp2}, an exactly solvable non-relativistic
potential problem were considered which provides a smooth transition between 
the Coulomb and the harmonic oscillator problems.

The aim of this paper is to extend this result for relativistic Coulomb
Green's operators, i.e., for the Coulomb Green's operator of the Klein-Gordon
and of the second order Dirac equations. This later is equivalent to the
conventional Dirac equation and seems to have several advantages.
For details see  Ref.\ \cite{hostler} and references therein.
The Coulomb Sturmian matrix elements of the second order Dirac equation 
has already been obtained by Hostler \cite{hostler} via evaluating complicated
contour integrals. Our derivation, however, is much simpler, it relies only
on the Jacobi-matrix structure of the Hamiltonian, and the 
result obtained is also better suited for numerical calculations. 
In Ref.\ \cite{hostler} the result appears in terms  of $\Gamma$ and
hypergeometric functions, while our procedure results in an easily  
computable and analytically continuable continued fraction.

\section{Matrix elements of relativistic Coulomb--Green's operators}
The radial Klein-Gordon and second order Dirac  equations for Coulomb
interaction are given by
\begin{equation}
\label{raddirac}
H_{u}  \left| \xi ^{u }\right\rangle =0, 
\end{equation}
where 
\begin{equation}
H_{u} = \left( \frac E{\hbar c}\right) ^2-\mu ^2+
\frac{2\alpha Z}{\hbar c}\frac Er+\frac{d^2}{dr^2}-\frac{u \left(
u +1\right) }{r^2}.
\end{equation}
Here $\mu=mc/\hbar$, $\alpha=\mbox{e}^2/\hbar c$, $m$ is the mass
and $Z$ denotes the charge.
For the Klein-Gordon case $u$ is given by
\begin{equation}
u =-\frac12 +\sqrt{\frac14+l(l+1)-(Z\alpha)^2},
\end{equation} 
and in the case of the second order Dirac equation
 for the different spin states we have
\begin{equation}
u_{\pm}= -\frac 12 \mp \frac 12 + 
\sqrt{\left( j+\frac 12\right) ^2- (Z\alpha)^2}.
\end{equation} 

The relativistic Coulomb Green's operator 
is defined as the inverse  of the  Hamiltonian $H_{u}$:
\begin{equation}
\label{greendef}
H_{u }G_{u }=G_{u }H_{u }={\mathbf{1}}_{u },
\end{equation}
where ${\mathbf{1}}_{u }$ denote the unit operator of the radial 
Hilbert space ${\cal H}_{u }$.

In complete analogy with the non-relativistic case we can define
the relativistic Coulomb Sturmian functions as solutions of 
the Sturm-Liouville problem
\begin{equation}
\label{sturmdef}
\left( -\frac{d^2}{dr^2}+\eta ^2+\frac{u \left( u +1\right) }
{r^2}-\frac{2\eta (n+u +1)}r\right) S_{n;\eta }^{u }(r)=0,
\end{equation}   
where  $\eta$ is a real parameter and $n=0,1,2, \ldots,\infty$ is
the radial quantum number. 
In coordinate space representation  they take the form,
\begin{equation}
\label{sturm}
  \left\langle r\left| n u;\eta \right.
\right\rangle =\left[ \frac{n!}{\left( n+2u+1\right) !}\right]
^{\frac 12}\left( 2\eta r\right) ^{u+1}e^{-\eta r}L_n^{2u+1}(2\eta r), 
\end{equation}
where $L$ is a  Laguerre-polinom.
The Coulomb Sturmian   functions, together with their  biorthogonal partner
$\left\langle r\right.
 \widetilde{\left| n u;\eta \right\rangle }%
=1/r\cdot \left\langle r\right. \left| n u;\eta \right\rangle $,
form a basis: i.e., they are orthogonal, 
\begin{equation}
\label{orto}
\widetilde{\left\langle n u;\eta \right. } \left| m u;\eta
\right\rangle =\left\langle n u;\eta \right. \widetilde{\left| m u
;\eta \right\rangle }=\delta _{nm}, 
\end{equation}
and form a complete set in ${\cal H}_{u}$
\begin{equation}
\label{completeness}
\sum_{n=0}^\infty {\left| n u;\eta \right\rangle }
\widetilde{\left\langle n u;\eta \right| }=\sum_{n=0}^\infty \left|
n u;\eta \right\rangle \widetilde{\left\langle n u;\eta \right| }
={\mathbf{1}}_{u}. 
\end{equation}

A straightforward calculation yields:
\begin{equation}
\label{overlap}
\begin{array}{l}
\displaystyle\vspace{3mm}\left\langle n u;\eta \right| \!\left.
m u;\eta \right\rangle \!\!=\!\!\frac 1{2\eta }\!\left[ \!\delta
_{nm}\!\left( 2u+2n+2\right) -\delta _{nm-1\!}\sqrt{\left( n+1\right)
\!\left( n+2u+2\right) }\right. \\ 
\displaystyle\hspace{3cm}\left. -\delta _{nm+1}\sqrt{n\left( 2u 
+n+1\right) }\right]. 
\end{array}
\end{equation}
Utilizing  this relation and considering Eq.\ (\ref{sturmdef}) we 
can easily calculate the Coulomb Sturmian matrix elements of $H_{u}$,
\begin{equation}
\label{diracmatrix2}
\displaystyle 
\begin{array}{l}
\vspace{3mm}\underline{H}_{nm}:=
\displaystyle\left\langle nu;\eta \right| 
H_{u}\left| m u;\eta \right\rangle = \\ 
\vspace{3mm}\displaystyle\hspace{2cm}+\delta _{nm}\left( \frac{2\alpha zE}{%
\hbar c}-2(u+n+1)\eta +2(u+n+1)
\frac{(E/ \hbar c ) ^2-\mu ^2+\eta ^2}{2\eta }%
\right) \\ 
\vspace{3mm}\displaystyle\hspace{2cm}-\delta _{nm-1}\left( \frac{
(E/ \hbar c ) ^2-\mu ^2+\eta ^2}{2\eta }\sqrt{(n+1)(n+2u+2)}\right) \\ 
\displaystyle\hspace{2cm}-\delta _{nm+1}\left( \frac{(E/ \hbar c ) ^2-
\mu ^2+\eta ^2}{2\eta 
}\sqrt{n(n+2u+1)}\right),
\end{array}
\end{equation}
which happens to possess a Jacobi-matrix structure. So, the theorem of  
Ref.\ \cite{klp1} is readily applicable here.

Let us consider the $\infty\times\infty$  Green's matrix
\begin{equation}
\label{greenmatrix}
\left( \underline{G}_{u}\right) _{nm}\equiv \widetilde{\left\langle
n u;\eta \right| }G_{u}\widetilde{\left| m u;\eta
\right\rangle },  
\end{equation}
and let us denote its rank-$N$ leading
principal submatrix by $\underline{G}_{u}^{(N)}$. 
Then, according to Ref.\ \cite{klp1},
\begin{equation}
\label{greeninverz}
(\underline{G}_{u}^{(N)})^{-1}_{i j}=\underline{H}_{ij}+
\delta _{j N}\ \delta _{i N}\ \underline{H}_{i N+1}\  {\cal F},  
\end{equation}
where  $\cal{F}$ is a  continued fraction
\begin{equation}
\label{cf}
 {\cal F}=-K_{i=N}^\infty 
 \left( \frac{a_i}{b_i}\right) =-\frac{a_{1+N}}{b_{1+N}}
{ \atopwithdelims.. +}
\frac{a_{2+N}}{b_{2+N}}
{ \atopwithdelims.. +\cdots +}
\frac{a_{n+N}}{b_{n+N}}
{ \atopwithdelims.. +\cdots },
\end{equation}
whose  coefficients   are related to 
the  Jacobi matrix 
\begin{equation}
\label{coefficients}
a_i  =  -\frac{\underline{H}_{i i-1}}{\underline{H}_{i i+1}},\ \ \  
b_i =  -\frac{\underline{H}_{ii}}{\underline{H}_{ii+1}}.
\end{equation}
This continued fraction convergent for bound-state energies, but, by using the
method presented in Ref.\ \cite{klp1}, can be
continued analytically to the whole complex energy plane. Simple matrix
inversion gives now the desired Green's matrix.

In Table I we demonstrate the numerical accuracy of method by evaluating the
ground and some highly excited sates of  
relativistic  hydrogen-like atoms, which, in fact, correspond to the poles
of the Dirac Coulomb Green's matrix. In particular, the
zeros of the determinant of (\ref{greeninverz}) were located.
It should be noted  that irrespective of the rank $N$
the zeros should provide the exact Dirac results. In  Table I
we have taken $2\times 2$ matrices. Indeed, the results of
this method, $E_{\text{cf}}$, agree with the exact one 
in all cases, practically up to the machine accuracy, allowing thus to
study the fine structure splitting.

\section{Summary}
In this short note we have presented a practical and easy-to-apply procedure
for calculating the Coulomb Sturmian matrix elements of the 
Coulomb Green's operator of the Klein-Gordon 
and of the second order Dirac equations. The method is relied only on the
Jacobi-matrix structure of the corresponding Hamiltonians and
results in a continued fraction which can be continued analytically to the
whole complex energy plane.

\acknowledgments
This work has been supported by the OTKA contracts No.\ T17298 and T026233.

\begin{table}[h]
\centering
\hspace{-1.6cm}. 
\begin{tabular}{|l|l|l|l|l|}
\hline
  & energy levels & $E_{\mbox{cf}}$ & $E_{\mbox{D}}$ & 
$E_{\mbox{S}}$ \\ \hline
hydrogen & $1$S$_{1/2}$ & $-0.5000066521$ & $-0.5000066521$ & $%
-0.5 $ \\ \cline{2-5}
$Z=1$ & $2$P$_{1/2}$ & $-0.1250020801$ & $-0.1250020801$ & $-0.125$ \\ 
\cline{2-5}
& $2$P$_{3/2}$ & $-0.1250004160 $ & $-0.1250004160 $ & $-0.125$ \\ 
\cline{2-5}
& $50$P$_{1/2}$ & $-0.0002000002 $ & $-0.0002000002 $ & $ 
-0.0002 $ \\ \cline{2-5}
& $50$P$_{3/2}$ & $-0.0002000001 $ & $-0.0002000001 $ & $ 
-0.0002 $ \\ \hline
uranium & $1$S$_{1/2}$ & $-4861.1483347 $ & $-4861.1483347 $ & $-4232$
\\ \cline{2-5}
& $100$D$_{3/2}$ & $-0.4241695002$ & $-0.4241695002$ & $-0.4232$ \\ 
\cline{2-5}
& $100$D$_{5/2}$ & $-0.4238303306$ & $-0.4238303306$ & $-0.4232$  \\
\hline
\end{tabular}

\caption{Energy levels of hydrogen-like atoms in atomic units.
$E_{\mbox{cf}}$ is the relativistic spectrum calculated via continued fraction,
 $E_{\mbox{ D }}$ and $E_{\mbox{\text S}}$ are 
textbook values of the relativistic Dirac
and the non-relativistic Schr\"odinger spectrum, respectively.}
\label{tablazat}
\end{table}

\end{document}